\def\be{\begin{equation}}
\def\ee{\end{equation}}
\def\ba{\begin{array}}
\def\ea{\end{array}}

\def\Cb{{\Bbb C}}

\documentclass[12pt]{article}
\input amssym.def

\newcommand{\lin}{\mathop{\rm span }\nolimits}
\begin{document}

\begin{center}
    {\large\bf  Classification of Bipartite and Tripartite Qutrit Entanglement under SLOCC}

    \vskip .6cm {\normalsize Xin-Gang  Yang$^{1}$,  Zhi-Xi Wang$^{1}$,  Xiao-Hong Wang$^{1}$  and  Shao-Ming  Fei$^{1,2}$}

    \vskip.4cm
\begin{minipage}{5.5in}

\small $~^{1}$ {\small College of Mathematical Scicences, Capital
Normal University, Beijing 100037, China}


{\small $~^{2}$ Max-Planck-Institute for Mathematics in the
Sciences, 04103 Leipzig, Germany}

\end{minipage}

\end{center}

\vskip 1.5mm

{\bf Abstract:~} We classify biqutrit and triqutrit pure states
under stochastic local operations and classical communication. By
investigating the right singular vector spaces of the coefficient
matrices of the states, we obtain explicitly two equivalent classes
of biqutrit states and twelve equivalent classes of triqutrit states
respectively.



\section{Introduction}

It is well known that quantum entanglement plays very important roles in
quantum information theory. One of the main tasks in quantum
information theory is to find out how many different ways
multipartite pure states can be entangled. It has been
shown that entangled qudits are less affected by noise than
entangled qubits \cite{2,3}. In quantum cryptography it is more secure against
evesdropping attacks using entangled qutrits or qudits than using qubits
\cite{4,5,6,7}. These facts motivate our
interest in multi-dimensional entangled states.

As the concept of entanglement is related to the nonlocal properties
of a state, local quantum operations can not affect the intrinsic
nature of entanglement\cite{8}. It is natural and meaningful to
classify pure states in terms of stochastic local operations and
classical communication (SLOCC). In \cite{9} it was shown that SLOCC
equivalent pure states can carry out the same quantum-informational
tasks with non-null possibly different possibilities, and two
$N$-partite states $\Psi$ and   $\Phi$  are equivalent under SLOCC
if and only if there exist invertible local operations (ILOs)
$F^{[1]}, \cdots, F^{[N]}$ such that  $\Psi=F^{[1]} \otimes F^{[2]}
\otimes \cdots \otimes F^{[N]} \Phi$.

In recent years, a lot of efforts have been made on classification
of multipartite entanglement under SLOCC [3, 9-18]. In \cite{1} an
inductive method of classifying $n$-qubit entanglement under SLOCC
has been presented, from which the entanglement classification of
three and four qubits have been obtained. In the classification of
biqutrit pure states some entanglement measures have been also used
in \cite{4}. In \cite{10} a range criterion has been used to judge
the equivalence of two states under SLOCC. The classification of
entanglement in $2\times m\times n$ systems is investigated. In
\cite{11} the complete SLOCC classification of multipartite
entanglement in $2\times 2\times n$ cases has been studied in two
different ways. It has been proved that a pure state of four qubits
can be transformed into nine families by SLOCC operations with
determinant one \cite{14}. In \cite{15} three-qubit states under
SLOCC on the basis of the canonical forms and on the local unitary
operator polynomial invariants have been classified.

In this paper we study the classification of biqutrit and triqutrit
pure states under SLOCC by using the method introduced in \cite{1}.
According to the dimensions of the right singular vector spaces of
the coefficient matrices of the states, we obtain explicitly
equivalent classifications of biqutrit states and triqutrit states
under SLOCC.

\section{Classification of biqutrit entanglement}

Let $\{e_{i}\}_{i=1,\dots,m}$ and $\{f_{j}\}_{j=1,\dots,n}$ denote
bases in $\Cb^{m}$ and $\Cb^{n}$, respectively. Any bipartite state
$\Psi\in C^{m}\otimes C^{n}$ can be written as
\begin{equation}\label{Psi}
\Psi=\sum_{i=1}^{m}\sum_{j=1}^{n}c_{ij}\,e_{i}\otimes f_{j},
\end{equation}
where $c_{ij}\in\Cb$. We denote $C=(c_{ij})_{m,n}$ the coefficient
matrix of the state $\Psi$.

According to the singular value decomposition, an $m\times n$ matrix
$C$ can always be decomposed as $C=V\Sigma W^{\dagger}$, where $V$
and $W$ are unitary matrices and $\Sigma$ is a diagonal matrix with
non-negative entries (singular values),
$\Sigma_{ij}=\sigma_{i}\delta_{ij}$, $i=1,\dots,m$, $j=1,\dots,n$
and $\sigma_{k}\geq 0$ for all $k$. The columns $v_i$ of $V=[v_{1}\
v_{2}\ \dots\ v_{m}]$ (resp. $w_i$ of \ $W=[w_{1}\ w_{2}\ \dots\
w_{n}]$) are called left (resp.\ right) singular vectors of $C$.

Since the relevant singular vectors will be those associated to
non-null singular values, in the following, we agree on referring as
singular vectors only to those $v_{k}$ and $w_{k}$ for which
$\sigma_{k}>0$. We denote by $\Gamma$ (resp.\ $\Pi$) the subspace
generated by the left (resp.\ right) singular vectors, i.e.
$\Gamma=\lin\{v_{1},\dots,v_{k}\}$ (resp. $\Pi=\lin\{w_{1},\dots,
w_{k}\}$). From the Schmidt decomposition of bipartite pure states,
the state (\ref{Psi}) is separable if and only if $\dim W =1$ (or
$\dim V =1$) \cite{1}.

Let $\Psi,\bar{\Psi}\in \Cb^{m}\otimes \Cb^{n}$ denote two bipartite
states related by SLOCC, i.e.
\begin{equation}\label{slocc}
\bar{\Psi}=F^{[1]}\otimes F^{[2]}(\Psi),
\end{equation}
where $F^{[1]}$ and $F^{[2]}$ are non-singular operators upon
$\Cb^{m}$ and $\Cb^{n}$, respectively. Similar to the two-qubit case
\cite{1}, in terms of singular value decomposition one can prove
that the corresponding coefficient matrices $C$ of ${\Psi}$,
$\bar{C}$ of $\bar{\Psi}$ in an arbitrary product basis are related
through
\begin{equation}
\bar{C}=(F^{[1]^{T}}V)\Sigma (F^{[2]\dagger}W)^{\dagger},
\end{equation}
i.e. if $v_{j}$ and $w_{j}$ are the left and
right singular vectors of the coefficient matrix $C$ respectively,
then the new left and right singular vectors with respect to
$\bar{C}$ will be $F^{[1]^{T}}(v_{j})$ and $F^{[2]\dagger}(w_{j})$
respectively.

For simplicity, we will write it as $\bar{\Psi}=F^{[1]^T}\otimes
F^{[2]^\dag}(\Psi)$ in stead of (\ref{slocc}). We first consider the
biqutrit ($n=m=3$) case. In this case, the coefficient matrix of an
arbitrary pure state in $\Cb^3\otimes\Cb^3$ can be expressed as
\[
C=\left(
\begin {array}{ccc}
C_{11}&C_{12}&C_{13}\\
C_{21}&C_{22}&C_{23}\\
C_{31}&C_{32}&C_{33}
\end{array}
\right).
\]
The dimensions of the right singular subspaces $\Pi$ could be  $1$,
$2$ or $3$.

If $\dim \Pi=1$, we can choose ILOs $F^{[1]}$ and $F^{[2]}$ such
that
\begin{eqnarray*}
F^{[1]}(v_{1})=\frac{1}{\sigma_{1}}e_{1},\ \ F^{[2]}(w_{1})=e_{1},
\end{eqnarray*}
where $\{e_1, e_2, e_3\}$ denote bases of $\Cb^3$. The new
coefficient matrix $\bar{C}$ is then of the form
\[
\bar{C}=\left(
\begin {array}{ccc}
1&0&0\\
0&0&0\\
0&0&0
\end{array}
\right),
\]
which correspond to the product state $\Psi_{0}=|00\rangle$
(we denote $e_1=|0\rangle$, $e_2=|1\rangle$ and $e_3=|2\rangle$ as usual
in the following).

In the case $\dim \Pi=2$, we choose ILOs $F^{[1]}$ and $F^{[2]}$
such that
\begin{eqnarray*}
F^{[1]}(v_{1})=\frac{1}{\sigma_{1}}e_{1},\ \
F^{[1]}(v_{2})=\frac{1}{\sigma_{2}}e_{2},\ \ F^{[2]}(w_{1})=e_{1},\
\ F^{[2]}(w_{2})=e_{2}.
\end{eqnarray*}
The new coefficient matrix will turn to be
\[
\bar{C}=\left(
\begin {array}{ccc}
1&0&0\\
0&1&0\\
0&0&0
\end{array}
\right),
\]
which correspond to the state $\Psi_{1}=|00\rangle+|11\rangle$.

For the case $\dim\Pi=3$, we choose ILOs $F^{[1]}$ and $F^{[2]}$
such that
\begin{eqnarray*}
F^{[1]}(v_{1})=\frac{1}{\sigma_{1}}e_{1}, \ \ F^{[1]}(v_{2})=
\frac{1}{\sigma_{2}}e_{2}, \ \ F^{[1]}(v_{3})=\frac{1}{\sigma_{3}}e_{3},\\[3mm]
F^{[2]}(w_{1})=e_{1}, \ \ F^{[2]}(w_{2})=e_{2}, \ \
F^{[2]}(w_{3})=e_{3}.
\end{eqnarray*}
Then $\bar{C}$ turns out to be a $3\times 3$ identity matrix, and
the corresponding state is
$\Psi_{2}=|00\rangle+|11\rangle+|22\rangle$.

Therefore,  biqutrit states can be entangled in two inequivalent
ways ($\Psi_{1}$ and $\Psi_{2}$) under SLOCC. While in \cite{4}
biqutrit entangled states are classified into three types:
\begin{eqnarray*}
|I\rangle=\frac{1}{\sqrt{2}}(|11\rangle+|00\rangle),\ \
|II\rangle=\frac{1}{\sqrt{3}}(|11\rangle+|00\rangle+|-1-1\rangle),\\
|III\rangle=\frac{1}{\sqrt{6}}(|11\rangle+|-1-1\rangle+|10\rangle+|01\rangle+|0-1\rangle+|-10\rangle).
\end{eqnarray*}
We find that the type $|II\rangle$ is in fact
equivalent to the type $|III\rangle$ under SLOCC:
$|III\rangle$ can written as
\begin{eqnarray*}|III\rangle=\frac{1}{\sqrt{3}}[
|1\rangle\otimes\frac{1}{\sqrt{2}}(|1\rangle+|0\rangle)
+|0\rangle\otimes\frac{1}{\sqrt{2}}(|1\rangle+|-1\rangle)
+|-1\rangle\otimes\frac{1}{\sqrt{2}}(|-1\rangle+|0\rangle)],
\end{eqnarray*}
by choosing the ILOs
\begin{eqnarray*}
F^{[1]}=\left(
\begin {array}{ccc}
1&0&0\\
0&1&0\\
0&0&1
\end{array}
\right),\ \
F^{[2]}=\frac{1}{\sqrt{2}}\left(
\begin {array}{ccc}
1&1&-1\\
1&-1&1\\
-1&1&1
\end{array}
\right),
\end{eqnarray*}
we get $F^{[1]}\otimes F^{[2]}|III\rangle=|II\rangle$.
This result can be also obtained by using the method of
Schmidt decomposition provided in \cite{9}.

\section{Classification of triqutrit entanglement}

An arbitrary triqutrit pure state $\Psi\in\Cb^3\otimes\Cb^3\otimes\Cb^3$ has the form
\begin{eqnarray}
\Psi=\sum_{i,j,k=1}^{3}c_{ijk}\,e_{i}\otimes e_{j}\otimes e_{k}.
\end{eqnarray}
We can write the coefficient matrix of $\Psi$ in the form
\[ C=C_{1|23}=\left(
\begin {array}{ccccccccc}
C_{111}&C_{112}&C_{113}&C_{121}&C_{122}&C_{123}&C_{131}&C_{132}&C_{133}\\
C_{211}&C_{212}&C_{213}&C_{221}&C_{222}&C_{223}&C_{231}&C_{232}&C_{233}\\
C_{311}&C_{312}&C_{313}&C_{321}&C_{322}&C_{323}&C_{331}&C_{332}&C_{333}
\end{array}
\right).
\]
There are also two other ways to write the coefficient matrix of
$\Psi$: $C_{2|13}$ and $C_{3|12}$. Without loss of generality, we
will use $C_{1|23}$ in below considerations.

The classification of triqutrit pure states is to choose the ILOs
$F^{[1]}$, $F^{[2]}$ and $F^{[3]}$ such that the final coefficient
matrix reduces to a canonical one. In order to do so, we have to
find all possible structures of the space $\Pi$.

Let $\Psi,\bar{\Psi}\in \Cb^{m}\otimes \Cb^{n}\otimes \Cb^{l}$
denote two tripartite states that are equivalent under SLOCC, i.e.
\begin{equation}
\bar{\Psi}=F^{[1]^T}\otimes (F^{[2]}\otimes
F^{[3]})^{\dagger}(\Psi),
\end{equation}
where $F^{[1]}$, $F^{[2]}$ and $F^{[3]}$ are non-singular operators
upon $\Cb^{m}$, $\Cb^{n}$ and $\Cb^{l}$, respectively. Similarly, we
can prove that the coefficient matrices $C$ and $\bar{C}$ in an
arbitrary product basis are related through
\begin{equation}
\bar{C}=(F^{[1]}V)\Sigma (F^{[2]}\otimes F^{[3]}W)^{\dagger}.
\end{equation}

Concerning the dimension of the right singular subspace $\Pi$, there
are again three possibilities($\dim\Pi=1$, $\dim \Pi=2$ and $\dim
\Pi=3$) for triqutrit case.

\subsection{The case of $\dim \Pi = 1$}

$1$. Case $\Pi=\lin\{\Psi_{0}\}$, where $\Psi_{0}$ is the product
state defined in the last section. In this case, $w_{1}$ is of the
form $w_{1}=\phi\otimes\psi$. We can choose ILOs $F^{[1]}$,
$F^{[2]}$ and  $F^{[3]}$ such that
\begin{eqnarray*}
F^{[1]}(v_{1})=\frac{1}{\sigma_{1}}e_{1},~~  F^{[2]}(\phi)=e_{1},~~
F^{[3]}(\psi)=e_{1}.
\end{eqnarray*}
Then the new coefficient matrix is of the form
\begin{eqnarray*}
\begin{array}{lll}
\bar{C}&=&\left(\begin{array}{ccc}\frac{1}{\sigma_{1}}&.&.\\0&.&.\\0&.&.\end{array}\right)
            \left(\begin{array}{ccccccccc}\sigma_{1}&0&0\\0&0&0&\\0&0&0\end{array}\right)
            \left(\begin{array}{ccccccccc}1&0&0&0&0&0&0&0&0\\.&.&.&.&.&.&.&.&.\\.&.&.&.&.&.&.&.&.\end{array}\right)\\[6mm]
&=&\left(\begin{array}{ccccccccc}1&0&0&0&0&0&0&0&0\\0&0&0&0&0&0&0&0&0\\0&0&0&0&0&0&0&0&0\end{array}\right),
\end{array}
\end{eqnarray*}
which corresponds to the state $e_{1}\otimes e_{1}\otimes
e_{1}\equiv|000\rangle$, and where the dots $\cdot$ indicates the
irrelevant character of that entry.

\noindent $2$. Case $\Pi=\lin\{\Psi_{1}\}$. In this case, the vector
$w_{1}=\phi_{1}\otimes\psi_{1}+\phi_{2}\otimes\psi_{2}$. Using the
same strategy as above, we obtain the state $e_{1}\otimes
e_{1}\otimes e_{1}+e_{1}\otimes e_{2}\otimes e_{2}$, which
corresponds to the canonical vector: $|000\rangle+|011\rangle$.

\noindent $3$. Case $\Pi=\lin\{\Psi_{2}\}$ = $\lin\{
\phi_{1}\otimes\psi_{1}+\phi_{2}\otimes\psi_{2}+\phi_{3}\otimes\psi_{3}\}$.
We obtain the state $e_{1}\otimes e_{1}\otimes e_{1}+e_{1}\otimes
e_{2}\otimes e_{2}+e_{1}\otimes e_{3}\otimes e_{3}$, which
corresponds to the canonical vector:
$|000\rangle+|011\rangle+|022\rangle$.

\subsection{The case of $\dim \Pi = 2$}

We first consider $\Pi=\lin\{\Psi_{0},\Psi_{0}\}$. One of the three
possible cases in this class is
$\Pi=\lin\{\phi\otimes\psi_{1},\phi\otimes\psi_{2}\}$. In this case,
$w_{1}=u_{11}\phi\otimes\psi_{1}+u_{12}\phi\otimes\psi_{2}$ and
$w_{2}=u_{21}\phi\otimes\psi_{1}+u_{22}\phi\otimes\psi_{2}$, where
the matrix $[u_{ij}]$ has rank two, since $w_{1}$ and $w_{2}$ are
linearly independent. We choose the ILOs $F^{[1]}, F^{[2]}$ and
$F^{[3]}$ such that
\begin{eqnarray*}
F^{[1]}_{1}(v_{1})=\frac{1}{\sigma_{1}}e_{1},\ \
F^{[1]}_{1}(v_{2})=\frac{1}{\sigma_{2}}e_{2},\ \
F_{2}^{[1]}=\left(\begin{array}{cc}\left(\begin{array}{cc}u_{11}^{*}&u_{12}^{*}\\
 u_{21}^{*}&u_{22}^{*}\end{array}\right)^{-1}&0\\0&1
 \end{array}\right),\\[3mm]
F^{[1]}=F_{2}^{[1]}F_{1}^{[1]},\ \ F^{[2]}(\phi)=e_{1},\ \
F^{[3]}(\psi_{1})=e_{1},\ \ F^{[3]}(\psi_{2})=e_{2}.
\end{eqnarray*}
The new coefficient matrix will be
\begin{eqnarray*}
\begin{array}{lll}
\bar{C}&=& F_{2}^{[1]}
            \left(\begin{array}{ccc}\frac{1}{\sigma_{1}}&0&.\\0&\frac{1}{\sigma_{2}}&.\\0&0&.\end{array}\right)
            \left(\begin{array}{ccccccccc}\sigma_{1}&0&0\\0&\sigma_{2}&0\\0&0&0\end{array}\right)
            \left(\begin{array}{ccccccccc}u_{11}^{*}&u_{12}^{*}&0&0&0&0&0&0&0\\
            u_{21}^{*}&u_{22}^{*}&0&0&0&0&0&0&0\\.&.&.&.&.&.&.&.&.\end{array}\right)\\[3mm]
&=&\left(\begin{array}{ccccccccc}1&0&0&0&0&0&0&0&0\\0&1&0&0&0&0&0&0&0\\0&0&0&0&0&0&0&0&0\end{array}\right),
\end{array}
\end{eqnarray*}
which corresponds to the state $|000\rangle+|101\rangle$.

Dealting with similarly, we can get two states
$|000\rangle+|110\rangle$ and $|000\rangle+|111\rangle$ from the
other cases($\Pi=\lin\{\phi_{1}\otimes\psi,\phi_{2}\otimes\psi\}$
and $\Pi=\lin\{\phi_{1}\otimes\psi_{1},\phi_{2}\otimes\psi_{2}\}$).

Using the same strategy, all together we obtain the following classifications:
\\

\begin{tabular}{|c|c|}\hline
{\rm\bf Class} & {\rm\bf Canonical Vector}\\\hline
$\lin\{\Psi_{0},\Psi_{0}\}$ & $|000\rangle+|101\rangle,\quad
|000\rangle+|110\rangle, \quad  |000\rangle+|111\rangle
$\\
\hline
$\lin\{\Psi_{1},\Psi_{1}\}$ & $\begin{array}{ll}
|000\rangle+|011\rangle+|101\rangle+|112\rangle,& |000\rangle+|011\rangle+|112\rangle+|120\rangle\\
|000\rangle+|011\rangle+|120\rangle+|101\rangle,&
|000\rangle+|011\rangle+|120\rangle+|102\rangle
\end{array}$\\\hline
$\lin\{\Psi_{0},\Psi_{1}\}$ & $\begin{array}{ll}
|000\rangle+|011\rangle+|101\rangle, & |000\rangle+|011\rangle+|112\rangle\\
|000\rangle+|011\rangle+|120\rangle, &
|000\rangle+|011\rangle+|122\rangle
\end{array}$\\\hline $\lin\{\Psi_{0},\Psi_{2}\}$ &
$|000\rangle+|011\rangle+|022\rangle+|101\rangle$\\\hline
$\lin\{\Psi_{1},\Psi_{2}\}$ & $\begin{array}{l}
|000\rangle+|011\rangle+|022\rangle+|101\rangle+|112\rangle\\|000\rangle+|011\rangle
+|022\rangle+|112\rangle+|120\rangle\\|000\rangle+|011\rangle+|022\rangle+|120\rangle+|101\rangle
\end{array}$\\\hline
\end{tabular}\\

We did not consider the case $\Pi=\lin\{\Psi_{2}, \Psi_{2}\}$. This
is due to that any two-dimensional subspace in $\Cb^{3}\otimes
\Cb^{3}$ contains at least one product vector $\Psi_{0}$ or one
entangled vector $\Psi_{1}$ with coefficient matrix rank two. This
can be got by the following way: Let $V$ be a two-dimensional
subspace of $\Cb^{3}\otimes \Cb^{3}$. Without loss of generality,
two entangled vectors of rank $3$ can be chosen as generators of $V$
with coefficient matrices given by $C_{1}=I$ and $C_{2}$ being an
arbitrary matrix of rank $3$ in the product canonical basis. Then it
is always possible to find non-null complex numbers $\alpha$ and
$\beta$ such that $\alpha I+\beta C_{2}$ has rank one or two:
because $-\beta/\alpha$ must be chosen to be an eigenvalues of
$C_{2}$, and if two eigenvalue of $C_{2}$ is the same, $\alpha
I+\beta C_{2}$ will have rank one.

\subsection{The case of $\dim \Pi = 3$}

We first consider $\Pi=\lin\{\Psi_{0},\Psi_{0},\Psi_{0}\}$. One of
the subcases is
$\Pi=\lin\{\phi\otimes\psi_{1},\phi\otimes\psi_{2},\phi\otimes\psi_{3}\}$.
In this case,
$w_{1}=u_{11}\phi\otimes\psi_{1}+u_{12}\phi\otimes\psi_{2}+u_{13}\phi\otimes\psi_{3}$,
$w_{2}=u_{21}\phi\otimes\psi_{1}+u_{22}\phi\otimes\psi_{2}+u_{23}\phi\otimes\psi_{3}$
and
$w_{3}=u_{31}\phi\otimes\psi_{1}+u_{32}\phi\otimes\psi_{2}+u_{33}\phi\otimes\psi_{3}$,
where the matrix $[u_{ij}]$ has rank three, since $w_{1},w_{2}$ and
$w_{3}$ are linearly independent. We choose the ILOs $F^{[1]},
F^{[2]} \ \ and \ \ F^{[3]}$ such that
\begin{eqnarray*}
F^{[1]}_{1}(v_{1})=\frac{1}{\sigma_{1}}e_{1},\ \
F^{[1]}_{1}(v_{2})=\frac{1}{\sigma_{2}}e_{2},\ \
F^{[1]}_{1}(v_{3})=\frac{1}{\sigma_{3}}e_{3},\ \
F^{[1]}_{2}=\left(\begin{array}{ccc}u_{11}^{*}&u_{12}^{*}&u_{13}^{*}\\
                                    u_{21}^{*}&u_{22}^{*}&u_{23}^{*}\\
                                    u_{31}^{*}&u_{32}^{*}&u_{33}^{*}\\
                                    \end{array}\right)^{-1},\\
F^{[1]}=F_{2}^{[1]}F_{1}^{[1]},\ \ F^{[2]}(\phi)=e_{1},\ \
F^{[3]}(\psi_{1})=e_{1},\ \ F^{[3]}(\psi_{2})=e_{2},\ \
F^{[3]}(\psi_{2})=e_{2}.
\end{eqnarray*}
Then the new coefficient matrix is
\begin{eqnarray*}
\begin{array}{lll}
\bar{C}&=& F^{[1]}_{2}
            \left(\begin{array}{ccc}\frac{1}{\sigma_{1}}&0&0\\0&\frac{1}{\sigma_{2}}&0\\0&0&\frac{1}{\sigma_{3}}\end{array}\right)
            \left(\begin{array}{ccccccccc}\sigma_{1}&0&0\\0&\sigma_{2}&0\\0&0&\sigma_{3}\end{array}\right)
            \left(\begin{array}{ccccccccc}u_{11}^{*}&u_{12}^{*}&u_{13}^{*}&0&0&0&0&0&0\\
            u_{21}^{*}&u_{22}^{*}&u_{23}^{*}&0&0&0&0&0&0\\u_{31}^{*}&u_{32}^{*}&u_{33}^{*}&0&0&0&0&0&0\end{array}\right)\\[3mm]
&=&\left(\begin{array}{ccccccccc}1&0&0&0&0&0&0&0&0\\0&1&0&0&0&0&0&0&0\\0&0&1&0&0&0&0&0&0\end{array}\right),
\end{array}
\end{eqnarray*}
which corresponds to the state  $|000\rangle+|101\rangle+|202\rangle$.

By investigating the rest cases similarly, all together we have the
following canonical states under SLOCC:

\begin{tabular}{|c|c|}\hline
{\rm\bf Class} & {\rm\bf Canonical Vector}\\\hline
$\lin\{\Psi_{0},\Psi_{0},\Psi_{0}\}$ & $\begin{array}{l}
|000\rangle+|101\rangle+|202\rangle,\quad
|000\rangle+|110\rangle+|220\rangle\\
|000\rangle+|111\rangle+|202\rangle,\quad
|000\rangle+|111\rangle+|220\rangle\\
|000\rangle+|111\rangle+|201\rangle,\quad
|000\rangle+|111\rangle+|222\rangle
\end{array}$\\\hline
$\lin\{\Psi_{0},\Psi_{0},\Psi_{1}\}$ & $
|000\rangle+|011\rangle+|1\phi\varphi\rangle+|2\chi\psi\rangle$\\\hline
$\lin\{\Psi_{0},\Psi_{0},\Psi_{2}\}$ & $\begin{array}{l}
|000\rangle+|011\rangle+|022\rangle+|101\rangle+|202\rangle\\
|000\rangle+|011\rangle+|022\rangle+|110\rangle+|220\rangle\\
|000\rangle+|011\rangle+|022\rangle+|101\rangle+|212\rangle
\end{array}$\\\hline
$\lin\{\Psi_{1},\Psi_{1},\Psi_{0}\}$ & $\begin{array}{l}
|000\rangle+|011\rangle+|101\rangle+|112\rangle+|2\phi\varphi\rangle\\
|000\rangle+|011\rangle+|112\rangle+|120\rangle+|2\phi\varphi\rangle\\
|000\rangle+|011\rangle+|120\rangle+|101\rangle+|2\phi\varphi\rangle
\end{array}$\\\hline
$\lin\{\Psi_{1},\Psi_{1},\Psi_{1}\}$ & $\begin{array}{l}
|000\rangle+|011\rangle+|101\rangle+|112\rangle+|202\rangle+|221\rangle\\
|000\rangle+|011\rangle+|101\rangle+|112\rangle+|210\rangle+|202\rangle\\
|000\rangle+|011\rangle+|101\rangle+|112\rangle+|221\rangle+|210\rangle\\
|000\rangle+|011\rangle+|112\rangle+|120\rangle+|202\rangle+|221\rangle\\
|000\rangle+|011\rangle+|112\rangle+|120\rangle+|221\rangle+|210\rangle\\
|000\rangle+|011\rangle+|120\rangle+|101\rangle+|221\rangle+|210\rangle
\end{array}$\\\hline
$\lin\{\Psi_{1},\Psi_{1},\Psi_{2}\}$ &
$\begin{array}{l}
|000\rangle+|011\rangle+|022\rangle+|101\rangle+|112\rangle+|202\rangle+|221\rangle\\
|000\rangle+|011\rangle+|022\rangle+|101\rangle+|112\rangle+|210\rangle+|202\rangle\\
|000\rangle+|011\rangle+|022\rangle+|101\rangle+|112\rangle+|221\rangle+|210\rangle
\end{array}$\\\hline
$\lin\{\Psi_{2},\Psi_{1},\Psi_{0}\}$ & $\begin{array}{l}
|000\rangle+|011\rangle+|022\rangle+|101\rangle+|112\rangle+|202\rangle\\
|000\rangle+|011\rangle+|022\rangle+|101\rangle+|112\rangle+|220\rangle\\
|000\rangle+|011\rangle+|022\rangle+|101\rangle+|112\rangle+|221\rangle
\end{array}$\\\hline
\end{tabular}\\
where $\phi,\,\varphi,\,\chi,\,\psi$ are pure states in $\Cb^3$.

According to structures of the space $\Pi$,
we have got that three qutrits
can be entangled in twelve inequivalent ways under SLOCC,
where the states
in the class of $\dim \Pi=1$, the first two states in $\lin\{\Psi_{0},\Psi_{0}\}$
and $\lin\{\Psi_{0},\Psi_{0},\Psi_{0}\}$ are either fully or bi-separable.

\section{Conclusion and Remarks}

We have shown that entangled states of two and three qutrits can be
classified into two and twelve equivalent types respectively under
stochastic local operations and classical communication, based upon
the analysis of the structure of the right singular subspace of the
coefficient matrix of the states in an arbitrary conical product
basis.

The range criterion \cite{10},  which can judge whether two pure
states are inequivalent under SLOCC, classifies multipartite
entanglement by analyzing the structure of the ranges of the states.
In fact, a closer study we can get that the range of a state is
equivalent to the right singular subspace of the state. As the ways
of entanglement are concerned, the result of theorem 2 in \cite{10}
is included in section 3.2.

The way of classifying pure states under SLOCC can be generalized to
high dimensional case by investigating the structures of the space
$\Pi$. It is easily seen that there are $n$ types of pure states in
the space $\Cb^{n}\otimes \Cb^{n}$ under SLOCC. For pure stats in
$\Cb^{n}\otimes \Cb^{n}\otimes \Cb^{n}$, one can also deal with
their classification according to the dimension of right singular
subspace. If the dimension of the subspace is 2, there will be
$C^{2}_{n}+(n-1)$ families of entanglement; If the dimension of the
subspace is 3, then $C^{3}_{n}+2(C^{2}_{n}-1)+(n-1)$ families; ...;
If the dimension of the subspace is $n-1$, then
$C^{n-1}_{n}+(n-2)(C^{n-2}_{n}-1)+...+2(C^{2}_{n}-1)+(n-1)$
families. All together the number of the classification is
$(n-1)^{2}+\sum_{i=2}^{n}[(1+i(n-i))C^{n-i}_{n}-i(n-i)]$, where
$C^{n-i}_{n}={n!}/{i! (n-i)!}$. For instance for $n=2$, we have
that, the result of \cite{9}, two qubits can be entangled in two
inequivalent ways.

\end{document}